# Frisch, Muller and Belot on an Inconsistency in Classical Electrodynamics[1]

Peter Vickers

ABSTRACT

This paper follows up a debate as to whether classical electrodynamics is inconsistent. Mathias Frisch makes the claim in *Inconsistency, Asymmetry and Non-Locality* ([2005]), but this has been quickly countered by Fred Muller ([2007]) and Gordon Belot ([2007]). Here I argue that both Muller and Belot fail to connect with the background assumptions which support Frisch's claim. Responding to Belot I explicate Frisch's position in more detail, before providing my own criticisms. Correcting Frisch's position, I find that I can present the theory in a way both authors can agree upon. Differences then manifest themselves purely within the reasoning methods employed.

**1** *Introduction*
**2** *Features of the theory*
**3** *Frisch's inconsistency claim*
**4** *Defending Frisch*
  **4.1** *Muller*
  **4.2** *Belot*
**5** *Difficulties for Frisch and a compromise*
**6** *Conclusion*

## 1 Introduction

Classical electrodynamics is the latest example in a long line of allegedly inconsistent theories, including Bohr's theory of the atom and Newtonian cosmology. Several authors take the existence of inconsistent theories as read, and go on to draw conclusions about rationality and the applicability of paraconsistent logics (Meheus [2002]). However, most if not all of the examples given are controversial. Usually the issue is not whether what is presented is inconsistent, but whether what is presented is *a theory*. The question, then, is how we can decide precisely what should constitute a given theory. Should one look to the textbooks and problem-solving techniques of the relevant community, to what that community believe to be true, or approximately true, or what? Is there, in fact, a single set of assumptions which constitute a given theory? I will attempt to answer some of these questions in this paper, by following up the recent debate as to whether classical electrodynamics is inconsistent.

---

[1] Paper forthcoming in the *British Journal for the Philosophy of Science*.

In §2 I give the basic outline of the theory, and introduce some important conceptual complications which accompany the Lorentz force equation. §3 takes us to Frisch's inconsistency claim, and how he manages to derive a contradiction. In §4 Frisch's claims are defended against the recent criticisms of Muller and Belot, before I provide my own criticisms in §5. The discussion leads us to a *consistent* construal of classical electrodynamics which I believe satisfies both of the conceptual frameworks in play. §6 is the conclusion.

## 2 Features of the theory

The theory in question is Maxwell-Lorentz classical electrodynamics (CED). It acts to explain electromagnetic phenomena by describing interactions between microscopic charged particles and electromagnetic (EM) fields. The ontological distinction between the particles and the fields divides the laws of the theory in two. On the one hand we have the Maxwell equations (MEs) which tell us how particles give rise to and affect fields. On the other hand we have the Lorentz force equation (LFE) which tells us how fields affect particles.

The link between the two parts of the theory is energy conservation. It is assumed that, for a given system of particles and fields, the sum of the kinetic energies $E_k$ of the particles and the field energy $E_f$ remains constant over time. Since energy is transferred between particles and fields, this means that for any increase in particle energies there is a corresponding decrease in field energy, and vice versa. For a closed system we can write,

$$\Delta E_k + \Delta E_f = 0.$$

If energy is leaving or entering by crossing an imaginary surface enclosing the system the conservation equation becomes,

$$\Delta E_k + \Delta E_f = E_{\text{over surface}}.$$

Strictly speaking this is meant to stand for any given system, and for any period of time however small.

The LFE will be central to this paper, so I should say a little more about it. As I said, it tells us how fields affect particles. It is often presented as

$$\mathbf{F} = q(\mathbf{E} + \mathbf{v} \times \mathbf{B}),$$

with the vector quantities $\mathbf{E}$ and $\mathbf{B}$ denoting the electric and magnetic field properties at the point (or points) of the particle, and q, $\mathbf{v}$ and $\mathbf{F}$ denoting, respectively, the charge on, velocity of, and (Lorentz) force experienced by the particle.[2] The idea is that we already know **E**, **B**, q and **v**, and we use the equation to determine **F**. However, there are some important complications hidden in the interpretation of **E** and **B**, and the terminology in the literature can be misleading.

Consider a single charged particle in uniform motion at time $t_0$. Which sources of field must the LFE take into account? The first and most obvious are the *external* fields, those which are reaching the particle at time $t_0$ from other charged particles. It is fundamental to CED that external fields affect the particle, so they should play a part in the LFE. To make things clear, these fields are sometimes distinguished from other fields by the subscript 'ext'. We can write,

$$\mathbf{F}_{ext} = q(\mathbf{E}_{ext} + \mathbf{v} \times \mathbf{B}_{ext}),$$

which tells us the force due purely to external fields. The total force is the combination of the external force and the so-called 'self-force': $\mathbf{F}_{tot} = \mathbf{F}_{ext} + \mathbf{F}_{self}$. It remains to determine what '$\mathbf{F}_{self}$' stands for. If external fields cause the external force, what are the *self*-fields which cause the *self*-force?

The most obvious self-field is the familiar Coulomb field. In its most common, introductory form it is a purely electric field which surrounds any charged particle, diffusing out into space isotropically and pointing directly away from the particle. It is proportional to the charge of the particle and inversely proportional to the square of the distance from it, so we can write,

$$\mathbf{E}_{coulomb} = k\frac{q\hat{\mathbf{x}}}{r^2}$$

(where $\hat{\mathbf{x}} = \mathbf{x}/|\mathbf{x}|$ and k is a constant). This field can be thought of as attached to the particle, as it accompanies it everywhere it goes. No energy is carried away from the particle; arrows pointing away from the particle which represent the field at a point don't represent the actual movement of anything, but just give the direction another, similarly charged particle *would* be pushed *if* one were placed there.[3] It acts on *other* particles as an *external* field, but here we are asking whether it should play any part in the LFE, where we consider the effect on the particle which plays host to that field.

We might now suppose that the particle must either be a point particle or be extended. If it is a point particle, then of course at the point of the particle r=0, so the Coulomb field there is going to be infinite in magnitude. Even for the extended case the field at the surface of the particle is going to be very large, since r will be very small. However we might find some comfort in the fact that, for a stationary particle (or indeed one in an inertial frame) the Coulomb field will surround the particle symmetrically. As such any forces will all cancel each other, to give zero net force. So in this case at least, the Coulomb field should *not* play any part in the LFE.

---

[2] Following Frisch and Belot I take it that I may ignore the relativistic generalisation of CED for the majority of the present discussion. However, see §§4.2 and 5, below.

[3] The flux over the surface of a sphere enclosing the particle is non-zero, and 'flux' comes from the latin for 'flow', so one could be forgiven for imagining something flowing away from the particle. However, 'flux' should only be taken to mean 'flow' when the vector field in question is representing movement. Compare the contours of a hill to the blowing of the wind.

However, when the particle is accelerating this symmetry breaks down. One might imagine the Coulomb field lagging behind its particle and constantly trying to catch it up. In this case it looks like the particle *will* experience a force due to its own field. But how can this be written into the LFE? The details will depend essentially on the structure of the particle, whether it is a point, a rigid sphere, a rigid shell, a dumbbell, a non-rigid body, or whatever. If the particle is not a point, then the difference between the field strength on one side of the body and the other is going to play a part, adding to the complications. So although a particle's own Coulomb field is relevant to the Lorentz force on a particle in certain circumstances, it isn't obvious how to include it in the LFE.

The term 'Coulomb field' is usually reserved for the isotropic, electric field which surrounds a charged particle when observed from its own rest frame. But when we consider the particle to be accelerating, and we introduce relativity, the field is in general neither purely electric nor symmetric. Authors often use the term 'static field' instead (Duffin [1990]; Jackson [1999]) for the field which is in some sense 'attached' to the particle (so the Coulomb field is a special case of a static field). The static field provides the contrast for another type of self-field, the 'radiation field'.

As is well known, when charged particles accelerate they emit radiation. In the classical theory this radiation is characterised as an electromagnetic wave, which falls off in intensity with 1/r from its host particle. It dominates the static field at large distances from the particle (since the latter goes with $1/r^2$), and is taken to exist independently of the static field. Thus it cannot be dismissed as a wave *in* a field; instead it takes on a life of its own, apparently independent of any medium. The things that 'wave' are electric and magnetic *field* properties which are directed perpendicular to the direction of travel and to each other. Since this radiation is caused by the particle, and is not a wave in a medium but carries its own fields with it, so to speak, it can fairly be described as emitted *field*. And it often is so described. Muller ([2007], p. 259) writes of

> The energy radiated by the moving charge, via its emitted electro-magnetic field, often called the *self-field*, $\mathbf{E}_{self}$ and $\mathbf{B}_{self}$.

Muller here apparently uses the term 'self-field' to refer to the radiation field alone, but it is much more usual to use this term to refer to *all* fields which are not external, radiation *and* static fields. Belot ([2007], p. 271) and Feynman ([1964],

p. 28-5f.) are cases in point, and I will follow their lead.[4] What is significant is that in all cases the emitted radiation is referred to as self-*field*. As Muller says, energy leaves the particle in the form of this radiation. Energy is conserved, the kinetic energy of the particle is decreased, and the particle's acceleration is reduced—so the particle does experience a self-force.

We have here another type of self-field affecting our particle, but again it's far from obvious how to write this into the LFE. For starters, the particle is taken to experience a force because it loses energy, not because it interacts with the radiation field's **E** and **B** components. If we do wish to include the radiated field in the LFE *as field*, then since it falls off with $1/r$ we experience essentially the same difficulties we saw with the $1/r^2$ Coulomb field when r becomes very small. How the field affects the particle will depend on the particle's structure.

However, it is in fact a fallacy to speak of the radiation field causing the host particle to experience a force. This attitude is prevalent in the literature; for example Griffiths ([1999]) writes, 'The radiation evidently exerts a force ($\mathbf{F}_{rad}$) back on the charge—a *recoil* force, rather like that of a bullet on a gun.' (p. 465). But the recoil on a gun really has the same, common cause as the ejection of the bullet, namely the rapidly expanding gas inside the chamber. Equally, the 'recoil' of a particle really has the same common cause as the emission of radiation. Feynman writes,

> With acceleration, if we look at the forces *between the various parts of the electron*, action and reaction are not exactly equal, and the electron exerts a force *on itself*. ([1964], p. 28-5f., former emphasis added)

In other words, the self-force is caused by the *static* fields of different parts of the electron affecting other parts as *external* fields, and when the particle accelerates these forces don't balance. And if the particle loses energy in this way, that energy has to *go* somewhere. The radiation field is the manifestation of this loss of energy. A little later in his book Griffiths tells a similar story (p. 472), and so too, briefly, does Belot ([2007], p. 269).

Where does all this leave us? Our characterisation of the LFE so far is $\mathbf{F}_{ext} = q(\mathbf{E}_{ext} + \mathbf{v} \times \mathbf{B}_{ext})$, plus an extra self-force effect when the particle is accelerating. We have two options for characterising this self-force:

(i)  Look to the energy emitted as radiation, and calculate the force on the particle using energy conservation;
(ii) Introduce a model of the particle, and calculate the force on different parts of the particle due to the static fields of other parts.

---

[4] In fact the distinction between 'static' and 'radiation' fields cannot be made in the relativistic generalisation of the theory, where one works with the field tensor *F* rather than the vector fields **E** and **B** (thanks to Mathias Frisch for emphasising this).

In fact, neither option leads to a satisfactory result. Griffiths and Jackson explore both possibilities. The first option (i) leads to the Abraham-Lorentz equation:

$$\mathbf{F}_{rad} = \frac{2q^2}{3c^3}\dot{\mathbf{a}}$$

(Griffiths [1999], p. 467; Jackson [1999], p. 748). Unfortunately this equation doesn't follow straightforwardly from the rest of the theory. Jackson notes that the derivation is "certainly not rigorous or fundamental" (p. 750). And in any case it carries with it severe conceptual difficulties. If it *did* follow rigorously from the rest of the theory the conceptual difficulties attached to it might be used to make a quite different inconsistency claim to the one in question in this paper.

The other option (ii) makes $\mathbf{F}_{self}$ a function of the particle's self-fields, $\mathbf{E}_{self}$ and $\mathbf{B}_{self}$, rather than its acceleration. It is usual, as noted above, to introduce a model of the particle in question and calculate how it interacts with its own fields. Now if we think of different parts of the particle affecting each other as *external* fields, then the equation for $\mathbf{F}_{self}$ will mirror the equation for $\mathbf{F}_{ext}$:

$$\mathbf{F}_{self} = q(\mathbf{E}_{self} + \mathbf{v} \times \mathbf{B}_{self}).$$

$\mathbf{F}_{self}$ is just the *external* force on a given *part* of the particle due to other parts of the particle. When we add to this the *other* external forces we get a 'total-LFE': $\mathbf{F}_{tot} = q(\mathbf{E}_{tot} + \mathbf{v} \times \mathbf{B}_{tot})$. However, what we wanted to know was how fields affect particles *as a whole*. To know the *net* force a particle exerts on itself we need to know the structure of the particle, because only then can we add up all the $\mathbf{F}_{self}$ terms for all the different parts. This is awkward, because any proposed model of an electron is going to be highly speculative. In addition we might wonder whether each part has an effect *on itself* (cf. Griffiths, p. 472).

It is also possible to work towards such an equation *without* introducing a particle model (see §4.2, below). But, as we will see, $\mathbf{F}_{self} = q(\mathbf{E}_{self} + \mathbf{v} \times \mathbf{B}_{self})$ only follows from the *relativistic* version of the theory, and is in any case of no *use* without introducing a model of the particle involved.

It is in trying to come to terms with these difficulties with the LFE, and how they are solved in practice, that Frisch's inconsistency claim is born.

### 3 Frisch's inconsistency claim

Frisch states that, '[T]he *core assumptions* of the Maxwell-Lorentz approach to microscopic particle-field interactions are inconsistent with each other.' ([2005], p. 34, emphasis added). He also calls these assumptions the 'fundamental principles' of CED (p. 39). He seems to be urging that he means to present a perfectly canonical construal of the theoretical content. *Prima facie* he is true to his word, presenting his assumptions on p. 33:

(1) There are in fact charged, accelerating particles.
(2) The MEs are valid.
(3) The LFE is valid.
(4) Energy is conserved in particle-field interactions.

However, in light of the discussion in the previous section we may ask just what precisely Frisch means by the LFE.

On p. 27 he introduces it as $\mathbf{F}_{Lorentz} = q(\mathbf{E}_{ext} + \mathbf{v} \times \mathbf{B}_{ext})$. Since only external fields are acting in this equation, it would seem to follow that $\mathbf{F}_{Lorentz}$ must stand for the force due only to external fields. This might then be explicitly written into the equation as $\mathbf{F}_{ext} = q(\mathbf{E}_{ext} + \mathbf{v} \times \mathbf{B}_{ext})$.

Much of what Frisch says is compatible with this interpretation of the LFE. He states (p. 30), 'The effect of *external* electromagnetic fields on charged particles is given by the Lorentz force law,' and (p. 35), '[T]he Lorentz force equation of motion *ignores* any effect that the self-field of a charge has on its motion.' (emphases added). However, it emerges that what he really means to say is, 'According to the Lorentz force law, the energy change of a charge is due *only* to the effects of external forces' (p. 33, emphasis added). It follows that he really construes the LFE as $\mathbf{F}_{tot} = q(\mathbf{E}_{ext} + \mathbf{v} \times \mathbf{B}_{ext})$, that is, the *total* EM force experienced by a charged particle is a function of the *external* fields only.

Now as noted in §2 we can write $\mathbf{F}_{tot} = \mathbf{F}_{ext} + \mathbf{F}_{self}$. Accordingly we can split Frisch's LFE into two parts, corresponding to the two different sources of fields at play. We get a purely external LFE, $\mathbf{F}_{ext} = q(\mathbf{E}_{ext} + \mathbf{v} \times \mathbf{B}_{ext})$, and a purely self-field LFE, $\mathbf{F}_{self} = 0$. It emerges that Frisch's LFE implicitly asserts that a charged particle experiences no force due to self-fields under any circumstances.

I said in the previous section that a particle *does* experience a force due to its self-field when it is accelerating, whether we imagine it interacting with its own static fields or imagine it recoiling from the emitted radiation field. In fact this follows from the MEs (2) and energy conservation (4): when a particle is accelerating the MEs tell us that energy is radiated, so by energy conservation the particle loses energy, and by work done it experiences a force, $\mathbf{F}_{self} \neq 0$. In other words we make the following inference:

$$\sim(E_{rad} = 0) \vdash \sim(\mathbf{F}_{self} = 0).$$

But we have just seen that Frisch's LFE tells us that in all circumstances (including when the particle is accelerating) $\mathbf{F}_{self} = 0$. We have here a contradiction following from Frisch's assumptions (1)-(4); thus (1)-(4) are inconsistent.

This isn't quite how Frisch presents the inconsistency, although it is closely related. The contradiction he derives is that the energy radiated by an accelerating point particle is both zero and non-zero: $E_{rad} = 0$ and $E_{rad} \neq 0$ (p. 34).

How can we show that $E_{rad} = 0$? Frisch argues as follows. Imagine a particle at position A at time $t_A$ undergoing a force which takes it to position B at time $t_B$. Then the kinetic energy $E_k$ of the particle at B will equal the kinetic energy of the particle at A plus the work done W by the force on the particle:

$$E_k(t_B) = E_k(t_A) + W \qquad (Eq.1)$$

However, according to Frisch's LFE, $\mathbf{F}_{tot} = q(\mathbf{E}_{ext} + \mathbf{v} \times \mathbf{B}_{ext})$, work is done on the particle only by the external fields $\mathbf{E}_{ext}$ and $\mathbf{B}_{ext}$. Thus Frisch writes that,

$$W = \int_A^B \mathbf{F}_{ext} \cdot d\mathbf{l}$$

He writes $\mathbf{F}_{ext}$, I submit, because from $\mathbf{F}_{self} = 0$ it follows that $\mathbf{F}_{ext} = \mathbf{F}_{tot}$. Therefore the work done W is meant to represent the *total* EM work done on the particle $W_{tot}$.[5] But now no energy can be radiated by the particle, since then we would have to write,

$$E_k(t_B) = E_k(t_A) + W - E_{rad} \qquad (Eq.2)$$

(Eq.1) is consistent with (Eq.2) only when $E_{rad} = 0$, but that $E_{rad} \neq 0$ follows from the MEs as I noted above.

In what follows I work with the contradiction $\mathbf{F}_{self} = 0$ & $\mathbf{F}_{self} \neq 0$, since this seems to clarify things a little. In fact the main difference between this and Frisch's presentation is as follows. I argue,

$$\sim(E_{rad} = 0) \vdash \sim(\mathbf{F}_{self} = 0).$$

Frisch argues,

$$(\mathbf{F}_{self} = 0) \vdash (E_{rad} = 0).$$

So I've really just focused not on Frisch's inference but on its contrapositive. The point of this manoeuvre lies in how clearly $\mathbf{F}_{self} = 0$ follows from Frisch's LFE. That $E_{rad} = 0$ is not so immediately clear.

At any rate, I agree that Frisch's construal of the theory is inconsistent, since a contradiction can be derived from it. So he *has* picked out a unit of analysis which is inconsistent. The obvious point of contention is his construal of the LFE, as he himself notes on p. 35:

---

[5] If Frisch intends W to represent the work done *due to external fields only* then no contradiction can be derived. Thus he probably ought to have written $\mathbf{F}_{tot}$ in his integrand rather than $\mathbf{F}_{ext}$ to make it clear that by 'W' he means the *total* work done.

> [T]he inconsistency is most plausibly seen as arising from the fact that the Lorentz force equation of motion ignores any effect that the self field of a charge has on its motion. The standard scheme treats charged particles as sources of fields and as being affected by fields—yet not by the total field, which includes a contribution from the charge itself, but only by the field external to the charge.

However, he has his reasons for taking his version of the LFE as canonical, as we shall see. In this passage he calls his construal of the theory 'the standard scheme'. We have already seen that he also refers to his assumptions (1)-(4) as the 'core assumptions' and the 'fundamental principles' of CED. This should be surprising, since we have already discussed in some detail how an accelerating particle *does* in fact experience a self-force. In what sense, then, can Frisch's construal be 'standard' or 'fundamental'?

These considerations suggest a natural fall-back position for Frisch if his claim that 'CED is inconsistent' doesn't hold up. He might claim, rather, that the unit of analysis he *has* introduced, although not the 'theory' of CED, is an interesting and important unit of analysis for philosophy of science. In what follows I intend to address both possibilities.

## 4 Defending Frisch

The community has been quick to criticise Frisch's claims, with Muller ([2007]) and Belot ([2007]) leading the way (all references to Muller and Belot will refer to these papers). In §5 I will indicate where I agree with these criticisms, and supply some of my own. But first I want to stress where I *don't* agree with the criticisms. Frisch's claims are grounded in a wider conceptual framework, and without carefully taking into account this framework it is all too easy to be dismissive. Apparent knock-down objections do not serve to undermine Frisch's conceptual foundations.

### 4.1 Muller

Muller's paper is extremely revealing for the debate, providing as it does a complete contrast of perspective from that put forward by Frisch. However, although it usefully fleshes out the intuitions of those already inclined towards Muller's point of view, it need not persuade those on Frisch's side of the fence.

The character of Muller's response is in evidence when in §1 he writes, 'we define the theory of CED in order to know exactly what is the object of Frisch's provocative charges.' He then goes on to give a *different* construal of the theory to that given by Frisch, since he introduces a different LFE. So in fact what Muller presents is *not* the 'object of Frisch's provocative charges', and he gives the impartial reader no reason to accept this version of the theory as *the* theory over Frisch's version.

With this as the starting point, Muller sees the case for the inconsistency of the theory in his own particular way. The argument should be that Frisch's inconsistency proof goes through fine, but that the 'thing', the 'unit of analysis' he finds to be inconsistent is not the theory of CED. Instead Muller writes that Frisch 'has applied CED inconsistently', and that 'the logic of the proof that has led him to this conclusion [is] flawed.' (§1). He goes on to say that Frisch makes 'two contradictory assumptions' such that 'we already have a contradiction by $\wedge$-introduction!' (p. 261). According to Muller, for an accelerating charged particle, Frisch first assumes $\mathbf{E}_{self} \neq 0 \neq \mathbf{B}_{self}$ and infers that $E_{rad} \neq 0$, then he assumes $\mathbf{E}_{self} = 0 = \mathbf{B}_{self}$ and infers that $E_{rad} = 0$. So, the story goes, Frisch assumes A and ~A, infers from A that B, and from ~A that ~B, and concludes that the theory is inconsistent since we have B&~B. Muller asks why he bothered to make the inferences, since he could have concluded A&~A at the start.

But this paints Frisch in a more confused light than is really fair. Muller seems to see *Frisch himself* as inconsistent, meaning that what he takes the theory to be is not held constant. But it is perfectly possible to interpret Frisch as he explicitly intends to be interpreted, as *consistently* employing the LFE $\mathbf{F}_{tot} = q(\mathbf{E}_{ext} + \mathbf{v} \times \mathbf{B}_{ext})$, and thereby presenting an inconsistent theory. From this perspective the weak point of Muller's argument is the claim that Frisch reasons from $\mathbf{E}_{self} = 0 = \mathbf{B}_{self}$ to $E_{rad} = 0$. We can infer from the discussion in §3, above, that Frisch argues from $\mathbf{F}_{self} = 0$ (via work done) to $E_{rad} = 0$. But from the fact that $\mathbf{F}_{self} = 0$ according to Frisch's LFE we cannot infer that $\mathbf{E}_{self} = 0 = \mathbf{B}_{self}$. At this stage we might have non-zero self-fields which just don't affect their host particle (cf. the quotation at the end of §3, above). Now if one considers the work done and draws on energy conservation, then one can conclude that $E_{rad} = 0$. But *still* we cannot conclude that $\mathbf{E}_{self} = 0 = \mathbf{B}_{self}$. At this stage we might have a self-field which carried no energy. One reaches the conclusion that $\mathbf{E}_{self} = 0 = \mathbf{B}_{self}$ by arguing, using the MEs, that an emitted EM wave will always carry energy, so that if no energy is being radiated there can be no EM wave emitted. So to get to $\mathbf{E}_{self} = 0 = \mathbf{B}_{self}$ in Frisch's theory we have to reason *from* $E_{rad} = 0$. In no sense does Frisch reason '$\mathbf{E}_{self} = 0 = \mathbf{B}_{self}$, therefore $E_{rad} = 0$'. He reaches $E_{rad} = 0$ by making inferences from his 'core assumptions' (1)-(4).

In answer, then, to the claim that CED is inconsistent Muller leaves us in no doubt that *for him* it is not. But we are lacking an argument as to why Muller's conception of the theory is canonical. And we are far from showing that Frisch's unit of analysis, even if not the canonical theory of CED, is not an interesting unit of analysis in its own right.[6]

---

[6] I would like to stress that Muller's paper comes highly recommended as a discussion of the difficulties faced by *his* particular construal of the theory. I just don't find it a fair appraisal of Frisch's position.

## 4.2 Belot

Belot makes a concerted effort to be more sensitive to Frisch's position. He tells us that the LFE comes in two different versions, one of which (Frisch's) is $\mathbf{F}_{tot} = q(\mathbf{E}_{ext} + \mathbf{v} \times \mathbf{B}_{ext})$. He presents the contradiction just as Frisch does, in terms of energy radiated, as $E_{rad} = 0$ & $\sim E_{rad} = 0$ (p. 271, fn. 11). The other version of the LFE Belot writes as $\mathbf{F}_{EM} = q(\mathbf{E} + \mathbf{v} \times \mathbf{B})$, and he soon makes it clear that this represents $\mathbf{F}_{tot} = q(\mathbf{E}_{tot} + \mathbf{v} \times \mathbf{B}_{tot})$. The purely self-field part of his LFE can then be written as $\mathbf{F}_{self} = q(\mathbf{E}_{self} + \mathbf{v} \times \mathbf{B}_{self})$.

As already noted in §2, stating the self-force in terms of self-fields is uninformative unless one provides a model of the particle in question. This Belot does, by introducing a rigid, spherical charged body with 'continuum many parts' (p. 269). The self-field is then taken into account by considering the force felt by each infinitesimal part of the rigid body by the other parts. Since these 'self-forces' are just special cases of external forces, Belot's self-LFE (and thus his total-LFE) takes the same form as the external-LFE. It can be shown that this secures consistency as follows.

Consider an accelerating charged particle within an imaginary sphere. From the discussion in §2 we know that for energy to be conserved we need the energy flowing over the surface of the sphere in a given time to equal the change in field energy within the surface in that time plus the change in particle energy within that time. In other words,

$$E_{\text{over surface}} = \Delta E_{\text{field}} + \Delta E_{\text{particle}} \qquad (\text{Eq.3})$$

Is this story reflected in the formalism?

First, working with the MEs one can take the surface integral of the Poynting vector to get the energy flow over the surface per unit time:

$$\frac{1}{\mu_0} \int_S \mathbf{E} \times \mathbf{B} \cdot d\mathbf{S}$$

(since $\mathbf{E}$ and $\mathbf{B}$ come from the MEs, they are the total fields). Now the MEs can be used to establish the following equality:

$$\frac{1}{\mu_0} \int_S \mathbf{E} \times \mathbf{B} \cdot d\mathbf{S} = -\frac{d}{dt} \int_V \frac{\mathbf{E} \cdot \mathbf{E} + \mathbf{B} \cdot \mathbf{B}}{8\pi} dV - \int_V \mathbf{J} \cdot \mathbf{E}_{tot} dV$$

The first term on the right hand side is the change in field energy inside S per unit time. So for energy to be conserved we want the second term on the right hand side to be the change in particle energy per unit time (cf. Eq.3).

Since we are dealing with a single particle instead of a current or charge distribution, we exchange the **J** in the final term for q**v** and do away with the volume integral. Also, it makes things easier if we multiply all terms by a tiny increment of time dt. This will allow us to talk of actual amounts of energy instead of energy per unit time. This then gives us,

$$\frac{dt}{\mu_0}\int_S \mathbf{E}\times\mathbf{B}\cdot d\mathbf{S} = -d\int_V \frac{\mathbf{E}\cdot\mathbf{E}+\mathbf{B}\cdot\mathbf{B}}{8\pi}dV - q\mathbf{v}\cdot\mathbf{E}_{tot}dt$$

Now the equality says that the energy flowing across the surface in time dt is equal to the change in field energy in that time minus q**v**.$\mathbf{E}_{tot}$dt. So if the latter expression gives us the change in particle energy we are on to a winner.

We get the change in particle energy from considering the LFE, and using the fact that the work done on the particle in time dt is equal to the change of kinetic energy of the particle in that time. Now the work done equals force times distance, W = **F**.d**l** for a tiny distance d**l**. And since we can write **v** = d**l**/dt we can also write the work done as **F**.**v**dt. Applying this to the LFE we get,

$$\mathbf{F}\cdot\mathbf{v}dt = q\mathbf{v}\cdot\mathbf{E}dt + q\mathbf{v}\cdot(\mathbf{v}\times\mathbf{B})dt = q\mathbf{v}\cdot\mathbf{E}dt$$

since $\mathbf{v}\cdot(\mathbf{v}\times\mathbf{B}) = 0$. So the change in energy of the particle in time dt is given by q**v**.**E**dt. So if the **E** in the LFE represents the *total* field, $\mathbf{E}_{tot}$, as it does in Belot's LFE, then energy is conserved. And of course if the **E** represents the external field only, $\mathbf{E}_{ext}$, as it does in Frisch's LFE, then energy is not conserved. So Belot's LFE takes into account the energy emitted by an accelerating charged particle, and thus turns Frisch's inconsistent CED into a consistent theory.

Belot now takes the crucial step, entering into a discussion of which version of the LFE is to be preferred (§VI). In what follows he gives voice to some of the *prima facie* advantages of his version over Frisch's:

(i)  '[Frisch's CED] does not deserve to be called a theory precisely because it is inconsistent.' (p. 277);
(ii) The total-LFE 'is fundamental', while the external-LFE 'is naturally seen as arising in the course of taking a useful approximation.' (p. 272);
(iii) '[T]he total-field version of the law should be preferred, as taking into account all of the actors involved.' (Ibid.);
(iv) '[E]nergy is not conserved in this [Frisch's] theory.' (Ibid.);
(v)  If you follow Frisch 'you can show that just about any theory is inconsistent.' (p. 275f.).

*Prima facie* these points are persuasive, but in fact they fail to do justice to the subtleties of Frisch's position. I take this opportunity to expand on Frisch's position, by responding to each in turn.

(i) In fact Belot doesn't explicitly use this as a reason to prefer his version of the LFE.[7] However, I want to maintain that the underlying motivation for the statement is the same as if he were to do so. Indeed, since at this stage he does intend to reject Frisch's version of CED as a *theory*, precisely because it is inconsistent, this in itself can count as a reason to prefer his version of events. This then stands as a good way to demonstrate an important difference between the two authors.

Belot's claim, as stated, is particularly strong. What he really means is as follows: 'Frisch's CED does not deserve to be called a theory precisely because it is *known* to be inconsistent.' There is then an obvious conception of 'theory' which would motivate this attitude: When we have a theory we have a *belief about the world*. The world is not inconsistent (or at least we don't *believe* it to be), so as soon as we know that a set of propositions are inconsistent they cannot constitute a theory.

An immediate rejoinder presents itself: what Frisch, Muller and Belot are discussing here is not CED as seen through the eyes of scientists at the beginning of the twentieth century. Rather, the focus of the enquiry is CED as it exists *today*. As such we do not believe it to be true, strictly speaking, since it has been superseded by quantum electrodynamics. But we still want to call CED a theory. So it can't be that it is necessary that a set of propositions are believed to be *true* for them to constitute a theory. But if we insist that we only believe in the *approximate* truth of a theory, then Belot's remark is unmotivated. Quite obviously approximately true propositions can be mutually inconsistent.

The obvious correction is that a theory must be a *possible* object of belief. That is, Frisch's CED is not a theory because it is not believ*able* (because it is known to be inconsistent), and the fact that we don't believe it (because it has been superseded) doesn't alter this. However, it isn't clear why this conception of theory should be preferred, particularly for superseded theories. Belot appeals to the 'fecundity' of insisting on consistency (p. 277), but this difficult term remains unexplicated.

However this is cashed out, it remains the case that Belot's motivation for his statement comes from a presupposition that what constitutes a theory is to be explicated in terms of belief. But Frisch need not be moved, since he explicitly states that his notion of theory acceptance does not depend on what is, was, or can be *believed*. He writes,

> [I]n accepting a theory, my commitment is only that the theory allows us to construct successful models of the phenomena in its domain, where part of what it is to be successful is that it represents the phenomena at issue to whatever degree of accuracy is appropriate in the case at issue. ([2005], p. 42)

---

[7] Thanks to Gordon for emphasising this.

Indeed, on this account it is not just that we should take a theory and adopt the type of commitment suggested. Rather, *the very content of a theory is defined by what is thus committed to*. I find this the only way to account for the way Frisch repeatedly identifies CED with '[T]he most common theoretical approach to modelling the interactions between charged particles and electromagnetic fields' (p. 1). More explicitly he writes, 'Throughout my discussion I will refer to the scheme used to model classical particle-field phenomena as a 'theory'.' (p. 26).

Admitting this conception of theoretical content then allows for inconsistent theories, and if theories are not allowed to be inconsistent Belot needs to tell us just what is wrong with this conception.

(ii) What does it mean to say that the total-LFE is 'fundamental'? Belot never broaches this difficult issue, and there is good reason to demand an analysis. The danger in the word is made manifest in that Belot uses it in two quite distinct ways. On p. 264 he describes CED as a 'less than fundamental theory'. In this context the word is apparently used to describe a theory which *has not* been superseded. However, on p. 272 he writes of the LFE, 'the total-field version of the law is fundamental while the external-field version is something that is naturally seen as arising in the course of making a useful approximation.' Now the word is used to describe a part of a theory which *has* been superseded.

One might attempt to unify the two different uses in terms of approximation. Non-fundamental theories give us approximate results, as Belot tells us on p. 280. There also seems to be a sense in which the external-LFE is an approximation, as Belot urges (p. 272). But, since CED is superseded even the *total*-field version only gives approximate results, and so is itself non-fundamental in the former sense. Does this dissolve the distinction Belot aims to establish?

One option is to talk of levels of fundamentalism. That is, the external-LFE is an approximation relative to the total-LFE, and the total-LFE is an approximation relative to a quantum-corrected LFE. So the quantum-LFE is more fundamental than the total-LFE, which in turn is more fundamental than the external-LFE. This preserves the distinction Belot is trying to draw. But what does it mean to say that the external-LFE is an approximation *relative to* the total-LFE? One thing it could mean is that the total version is truer. However, since it is strictly speaking false, this option requires delving into the well-known difficulties of approximate truth. I doubt Belot would want his argument to depend on that.

Another thing it could mean is that the total version gives better, more accurate predictions. However, this just isn't the case. In fact, as I will discuss shortly, Belot's version of the LFE never does give better models of the phenomena, or more accurate predictions. As I have already explained, it can't be used without introducing a model of the electron. But crucially when we *do* add such a model, the effects of the self-field are (virtually[8]) always negligible in the classical domain.

One final thing it could mean is that if the rest of CED *were* fundamental (not superseded), then the total-LFE would be, because *it follows* from the rest of CED. In fact it doesn't follow in a non-relativistic setting, as we will see. But even in a relativistic setting it's not clear that Frisch should be motivated by what *follows*. Indeed he *cannot* be, since his theory is inconsistent, and by *ECQ*[9] from an inconsistency *everything* follows. Crucially, he has a principled reason for not accepting what follows in the way Belot does. Belot is motivated in his construal of the theory, to some degree at least, by belief and truth. He is correspondingly motivated by *truth*-preserving inferences. But since Frisch's construal of the theory is based on model-building he need not be similarly motivated, since a truth-preserving inference need not be 'good-for-model-building'-preserving. This explains why, for Frisch, 'theories do not have a tight deductive structure' (p. 11).

What we find, then, is that Belot has told us why *he* is motivated to label the total-LFE fundamental and the external-LFE an approximation. He hasn't told us why Frisch should be similarly motivated.[10]

(iii) What does Belot mean by the LFE taking into account 'all of the actors involved'? At first this seems pretty clear cut. In Frisch's theory the MEs tell us that accelerating charged particles radiate energy. By energy conservation this will cause the particles to lose energy, and from this loss of energy we can infer a force on the particle. Therefore, it would seem, the self-fields should play a part in the LFE; they are 'involved'.

How does this look from Frisch's point of view? For starters, Belot needs to explain why he has taken the MEs and energy conservation and told us what Frisch's *LFE* should be like. Why hasn't he taken the MEs and the LFE and told us how energy conservation should change, or taken the LFE and energy conservation and told us how the MEs should change? And at any rate why should Frisch accept any one of these three options? Aren't we just demanding consistency once again?

We avoid begging the question by taking 'involved' to mean that self-fields really do affect charged accelerating particles in relevant experiments. This would be to say that, in the domain of classical EM phenomena, some of the phenomena are readily explained by the total-LFE but not by the external-LFE. However, Belot apparently would admit that there are no such phenomena.

---

[8] I will return to this important qualification in §5.

[9] *Ex contradictione quodlibet*: From 'A&~A' infer 'A' and '~A'. From 'A' infer 'AvB' for *any* arbitrary 'B'. From '~A' and 'AvB' infer 'B'.

[10] I interpret Muller's distinction ([2007], p. 261) between '=' and '≈' as making essentially the same point as Belot here. '=' corresponds to Belot's 'fundamental' and '≈' to what is 'approximate'.

Drawing on Jackson he tells us that the external-LFE is 'empirically adequate down to the level at which quantum effects begin to appear.' (p. 274). Frisch tells us the same thing: quantum effects become important before self-radiation effects do (p. 42).[11] Now, everyone agrees that CED need not concern itself with quantum effects—these are insignificant in the relevant phenomenological domain. But if self-force effects are equally negligible, then we must ask why these *must* feature in the theory. There is clearly a significant sense in which they are *not* 'involved', just as quantum effects are not.

If this is the case it may be urged that the problem then lies elsewhere in the theory. Perhaps one should take Frisch's external-LFE and energy conservation, and change the MEs to suit. Self-fields should be rejected altogether since they are apparently insignificant in the domain. However, of course it is only the *effects of self-fields on their host particle* which are insignificant. The self-fields themselves are perfectly significant, and were being readily 'observed' and manipulated in experiments as early as the 1890s, X-rays being the first important manifestation. So although self-*forces* are negligible within the classical domain, self-*fields* are not.

Therefore if we interpret Belot's 'being involved' as 'being significant' we see a possible motivation for the inconsistent construal of the theory. Frisch might argue as follows. Self-radiation comes under the effects of particles on fields, and therefore is the business of the MEs. This is non-negligible (in non-quantum contexts), and thus the MEs ought to include it. By contrast the self-*force* comes under the effects of fields on particles, and therefore is the business of the LFE. This *is* negligible, and thus the LFE ought *not* to include it (just as it doesn't include quantum considerations). A consistent account of particle-field interactions can't accompany such an emphasis on significance. That this underlies Frisch's account is suggested (p. 16) when he draws on Rohrlich and Harding. They cite electrodynamics as an example of an 'established theory', which is mature and successful and 'a permanent part of science', but which has been superseded ([1983], p. 603). The boundary of such a theory is defined according to its validity limits, which could be interpreted as placing the self-force outside the theory and the self-field inside.[12]

It seems that all Belot is left with is the claim that the self-fields are 'involved' because it follows from the rest of the theory that they are. However, as we have already seen, Frisch need not be moved on these grounds unless what follows is good for model-building. And he doesn't need the influence of the self-fields in his LFE, since he can produce all the representational models he needs for the classical domain with the external-LFE.

---

[11] We will see in §5 that this isn't *always* the case.

[12] I think this would be an abuse of Rohrlich and Harding's ideas, though, as they take their validity limits from the succeeding theory, and need not conclude that the self-force lies outside the theory given cumulative effects. See below.

(iv) That energy is not conserved in Frisch's account is repeatedly stated by Belot. However, Frisch has included energy conservation in his set of 'fundamental principles' of the theory, (1)-(4). It appears to me that all Belot is really saying here is that energy conservation is inconsistent with the other assumptions in Frisch's theory, and that this just won't do. So this is really just another way of saying that Frisch's CED is inconsistent, and 'something's gotta give'. Of course really energy conservation is no more violated in Frisch's scheme than are the MEs (2), or his LFE (3), or even the ontological claims (1).[13] But by Frisch's lights since all of the assumptions are used in different circumstances, all of them deserve a place in the theory. They just can't all be used together. As Frisch writes,

> [F]or a given system we use only a proper subset of the theory's equations to model its behavior, where the choice of equations *depends on what aspect of the interaction between charges and fields we are interested in.* (p. 40, original emphasis)

Here, once again, Belot is biased by his underlying conceptions of *what theories are* and *how they are used*. The point is that he hasn't argued for *these* conceptions over Frisch's (more on this in §5).

(v) According to Belot, Frisch will find that 'any reasonably complex physical theory is inconsistent.' (p. 275). This is because, as Belot sees it, Frisch is pursuing 'a desire to be faithful to the practice of physicists.' (p. 273). And, when we look at the practice, different parts of theories are ignored at different times for the sake of simplification and approximation. According to Belot, Frisch will then include in his theory a statement and its negation, since the negation will represent theoretical contexts where that particular theory-element is ignored.

But I think Belot has missed an important motivation for Frisch's content selection, already noted in section (iii). Frisch can argue that CED is inconsistent because the self-force is *always insignificant in the relevant domain*, even though the self-fields are not. Belot seems to be suggesting that for Frisch a theory will include a categorical statement just because some theoretical feature is only *sometimes* ignored. Frisch can reply that such a feature (e.g. $\mathbf{F}_{self}$) must *always* be ignored for the statement ($\mathbf{F}_{self} = 0$) to be included.

Belot cites as the main problem for Frisch here the fact that fundamental theories will end up inconsistent (p. 276, fn. 16); Frisch himself writes that fundamental theories cannot be inconsistent (p. 43). However, on my understanding of Frisch's position the motivation to leave the self-force out comes from considering the limit of the domain of CED—a limit that a fundamental theory would not have. It may turn out that other *superseded* theories will end up inconsistent, but certainly it's less obvious than if Frisch delineated theory content in the way Belot suggests. It would remain for Belot to provide examples of such theories, and at any rate I doubt this would worry Frisch very much.

---

[13] See Frisch, p. 51ff., for a discussion of how different (unappealing) ontological commitments render the theory consistent.

So Belot's criticisms of Frisch's position don't quite hit the mark, or at least don't give the requisite detail to do the intended damage. The recurring oversight on Belot's part is that Frisch's overall philosophy of science is not taken into account; his position is not assessed on its own terms. Indeed, inconsistencies in scientific theories may be a real difficulty if one's notion of theory acceptance rests on belief and truth, or if one insists on the deductive closure of theories (as Belot seems to). But Frisch does not sign up to either of these, so to criticise his account on the grounds that the latter must be rejected is no criticism at all. And not only do Belot's criticisms fall wide of the mark, but *prima facie* there are criticisms which might be made of Belot's own conception of the theory. I will now suggest that *his* version of the LFE is (B1) not useful, (B2) not significant, and (B3) doesn't follow from the rest of the theory.

(B1) As already noted, the self-field part of Belot's LFE—$\mathbf{F}_{self} = q(\mathbf{E}_{self} + \mathbf{v} \times \mathbf{B}_{self})$—requires a model of a charged particle if it is going to be useful, and Belot's paper attests to this when he introduces such a model. Upon deriving a similar equation, Jackson remarks, 'To calculate the self-force … it is necessary to have a model of the charged-particle.' (p. 751). Where does such a model come from? Belot would surely say from *outside* the theory. He writes,

> [T]he Maxwell-Lorentz equations no more pick out a structure for microscopic charged particles than Newton's law of gravity picks out a structure for microscopic massive particles: one is free to stipulate a notion of particle for a given investigation. (p. 266)

So in other words, by Belot's own admission it would seem, the self-force part of his LFE does no work within the theory; it takes the addition of auxiliary assumptions for it to become useful. But then perhaps *it* should also be an auxiliary assumption.

(B2) As discussed above in section (iii), even when we add a particle model to Belot's LFE it still isn't significant enough to be of interest in the domain of classical EM phenomena. So, starting any problem of CED with the two halves of Belot's LFE, $\mathbf{F}_{ext} = q(\mathbf{E}_{ext} + \mathbf{v} \times \mathbf{B}_{ext})$ and $\mathbf{F}_{self} = q(\mathbf{E}_{self} + \mathbf{v} \times \mathbf{B}_{self})$, one would immediately ignore the second half and work exclusively with the first. The suspicion arises that Belot's $\mathbf{F}_{self} = q(\mathbf{E}_{self} + \mathbf{v} \times \mathbf{B}_{self})$ really does stand just to ensure consistency.

(B3) As already noted, for Belot the content of a theory seems to be defined by the doxastic commitments of the relevant community.[14] Now despite (B1) and (B2) Belot might argue that his self-LFE should stand, because if one believes a set of propositions are true, and then uses truth-preserving inferences to reach another proposition, one should believe that proposition to also be true. However, it might be objected that these considerations do not put Belot's LFE in the theory either.

Recall from the beginning of this section that Belot's LFE was shown to be consistent with the MEs and energy conservation. What was required was for the following equality to hold:

$$\mathbf{F}_{tot} \cdot \mathbf{v} dt = q\mathbf{v} \cdot \mathbf{E}_{tot} dt$$

Belot's LFE succeeded here where Frisch's failed. Now we don't want to meddle with the external part of the LFE (it *is* significant, useful and well-confirmed), so what is important here is the following equality:

$$\mathbf{F}_{self} \cdot \mathbf{v} dt = q\mathbf{v} \cdot \mathbf{E}_{self} dt$$

This actually gives us quite a bit of freedom vis-à-vis the self-field part of the LFE. Instead of Belot's $\mathbf{F}_{self} = q(\mathbf{E}_{self} + \mathbf{v} \times \mathbf{B}_{self})$ we could just have $\mathbf{F}_{self} = q(\mathbf{E}_{self})$, and claim that the particle only experiences a force due to *electric* self-fields. More generally, the following equation satisfies the equality, where *g* stands for *any* 3-vector to 3-vector function:

$$\mathbf{F}_{self} = q\mathbf{E}_{self} + \mathbf{v} \times g(\mathbf{B}_{self}) \qquad (Eq.4)$$

It doesn't matter what vector we cross $\mathbf{v}$ with in the final term, $\mathbf{v} \cdot (\mathbf{v} \times \mathbf{A}) = 0$ for all $\mathbf{A}$.

In fact if we are working with the relativistic version of the theory Belot's self-LFE *would* follow from the rest of the theory. This is because the only way to make (Eq.4) covariant is to set $g(\mathbf{B}_{self}) = q\mathbf{B}_{self}$.[15] This would quash all three objections (B1)-(B3) in one fell swoop. However, Belot explicitly states that he is not considering the relativistic version of the theory. He writes (p. 266),

> [M]y present concern is with the question whether (1)-(5) are consistent—and special relativistic considerations form no part of (1)-(5).

---

[14] In fact Belot never gives us any criteria for theory membership, but I take doxastic commitment, of some sort or another, as underlying objection (i), above, and several other passages in his paper. See, for example, pp. 280-1.

[15] To show this, consider a moving charged particle from the perspective of two different frames of reference, relativistically transform $\mathbf{F}$, $\mathbf{E}$, $\mathbf{B}$ and $\mathbf{v}$, and consider what form *g* must take for (Eq.4) to be covariant.

In a non-relativistic setting Belot might try to justify his version of the $\mathbf{F}_{self}$ equation by arguing that the self-force is just a special case of the external force, as noted above, and that the two equations must therefore take the same form. But this would require particles to be extended, and Belot has said that the theory shouldn't take a stand on whether particles are extended or points (p. 266f.). In addition, there seems to be some tension between the external-LFE and the self-LFE, since the former refers to an effect on the particle *as a whole* (acting at the point of the centre of mass of the particle), whereas the latter refers to an effect on different *parts* of a particle. Finally, Belot must apparently reject the suggestion that each part of a particle affects *itself*, contra Griffiths ([1999], p. 472).

Objection (B3) throws up another apparent difficulty for Belot's program. If the content of a particular scientific theory is going to be decided by the doxastic commitments of a community, then it looks like we are going to have some difficult borderline cases. Whether we opt for belief in the truth or approximate truth of various assumptions, that belief is going to be a matter of degree. Often it is debatable whether an inference is truth-preserving, for example, and assumptions are going to be more or less-well confirmed. The upshot is that Belot's theory is liable to have fuzzy edges. A Lakatosian picture, with a sharp division between the hard core (theory) and the auxiliary assumptions, will not be representative.

Finally, not only does Belot have some work to do to justify his version of events, but he freely admits some advantages (were other things equal) of Frisch's conceptual scheme. Belot writes, 'The big advantage of this version [the external-LFE] is that it allows one to work with much simpler equations.' (p. 272) and, noting that Jackson explicitly adopts the external-LFE, he writes, 'So it appears that at the level of official doctrine and at the level of problem-solving, the external version of the Lorentz force law is taken as standard by physicists.' (p. 273). He also notes of Feynman, 'Much that he says gives the impression that he identifies classical electrodynamics with [Frisch's version of the theory].' (p. 274). Belot is keen to point out that both authors, Jackson and Feynman, do 'eventually' note that they have neglected radiation reaction. But it is clear that he does see the appeal of focusing on a set of assumptions which are central to 99% of the literature.

To sum up this long section, then, Belot's criticisms of Frisch largely fail to make their mark, and there are some difficulties for Belot's own construal of the theory which he doesn't address. Some important questions then remain. Can Frisch's position be criticised, or is it in some sense immune to criticism? Can we get to the bottom of Frisch's position and target its foundations? And can Belot's position be saved from its difficulties? Indeed, do we have a genuine conflict here, or are Frisch and Belot merely engaged in two different, mutually compatible analyses? I will address such questions in the next section.

## 5 Difficulties for Frisch and a compromise

We have now seen just how different Frisch's conception of CED is from Belot's. The difference in the construal of the LFE is only the beginning. Crucially, different conceptions of what deserves to be called a part of a theory accompany each LFE. On the one hand Belot looks, at least in part, to the doxastic commitments of the relevant community. Frisch, on the other hand, identifies CED with '[T]he most common theoretical approach to modelling the interactions between charged particles and electromagnetic fields.' (p. 1). Since each party insists on calling their unit of analysis 'the theory' there is little wonder miscommunication arises. But apart from this terminological disagreement, it looks like both research programs can exist side by side. Dividing the two units of analysis into 'theory$_B$' for Belot, and 'theory$_F$' for Frisch, we can happily say that 'theory$_B$' is consistent and 'theory$_F$' is inconsistent.[16]

What is then required to do justice to either party is to consider their conceptual scheme taken as a whole, on its own terms. We have seen that Belot makes the mistake of considering just one part of Frisch's overall conceptual scheme by the lights of his *own* scheme. Frisch at least considers the possibility of compatibility. He writes, 'a certain amount of peaceful coexistence between the two rival views of theories is possible, if we realize that they might be talking about different aspects of scientific theorizing.' (p. 12).[17] However, he insists that 'Genuine disagreements exist' (p. 11), and CED is meant to represent a situation where Belot's view fails and the models-based view succeeds. His argument is that, on Belot's view, 'accepting an inconsistent theory entails being committed to inconsistent sets of consequences.' (p. 41). And since CED *is* inconsistent, Belot *is* committed to inconsistent sets of consequences. Thus, 'Inconsistent theories [like CED] may be taken to provide particularly strong support for the importance of 'model-based' accounts of theories.' (p. 12).

However, as we have already seen, Frisch also argues *from* the models-based account of theory acceptance *to* the inconsistency of CED. But he can't have it both ways! All he has established here is the biconditional:

---

[16] One might draw here on Kenat ([1987], p. 87) who (himself drawing on a paper by Sylvain Bromberger) distinguishes two types of theory. 'Theories1' are 'theories as techniques for developing answers to problems', and 'theories2' are 'propositions'. See also (Suppe [1989], chapter 14).

[17] He goes on, "Sometimes scientists do seem to be interested in global representations of ways the world might be, and in such a case a possible worlds account of theories may well be philosophically illuminating." (Ibid.). I take it that this is representative of Belot's position.

Frisch's version of theory-acceptance ↔ CED is inconsistent.

If CED is inconsistent we need Frisch's account of theory-acceptance (or something like it). And if we employ Frisch's account of theory-acceptance, then since the content of a theory is defined by what the relevant community *accepts*, CED is inconsistent. So this just testifies to the internal coherence of Frisch's own conceptual scheme, and doesn't impinge on Belot's at all.

We might well ask at this stage whether (B1), (B2) and (B3) then do justice to Belot's conceptual scheme, and thus count as genuine criticisms. (B3) seems to, since it looks like Belot's self-LFE couldn't be a justified theory-element unless it followed from the rest of his theory. However, as I already noted, if Belot takes himself to be assessing the *relativistic* version of the theory, his LFE *does* follow. (B1) and (B2) then dissolve, since on Belot's terms it doesn't matter that the equation isn't useful or significant if it follows.[18]

In this way Belot's position looks fairly robust, but as we found fault with his criticism of Frisch's position we can't yet conclude that Belot's position is *better*. However, I now identify two new difficulties for Frisch which are sensitive to his conceptual framework: (F1) the domain of CED, and (F2) the role of $\mathbf{F}_{self} = 0$ in Frisch's theory.

(F1) In §4.2, where I was defending Frisch's position against Belot's criticisms, I drew on the 'fact' that the self-force is always insignificant in the domain of CED. In section (iii) we saw that Belot and Frisch both draw on Jackson to establish that the external-LFE is 'empirically adequate down to the level at which quantum effects begin to appear.' (Belot, p. 274). The relevant passage in Jackson is as follows: 'Only for phenomena involving such distances [$10^{-15}$m] or times [$10^{-24}$s] will we expect radiative effects to play a *crucial* role.' (p. 747, original emphasis). However, Jackson doesn't actually mean 'only for' here, for two reasons.

On the one hand the microscopic time and distance noted in the quotation given are derived from the Larmor formula for the radiation power of an accelerating charged particle. But Frisch is keen to stress that the theory he *really* takes himself to be analysing is *relativistic* CED. He writes, 'It [CED] is a *classical* theory only in that it is not a *quantum* theory.' (p. 29). And when one takes into account the relativistic power radiation formula, one finds that the radiation emitted depends on $\gamma^6$ (Jackson, p. 666). So for a particle travelling at speeds close to the speed of light, the power radiated increases dramatically, and thus the self-force becomes much more significant.

This still leaves us with fairly small times and distances, but there is another factor to take into account—the 'long-term, cumulative effects' (Jackson, p. 747). For example, since a particle in a synchrotron accelerator is continuously radiating the effects mount up. In early synchrotrons the loss per turn was only 1000 electronvolts, and yet even this was non-negligible compared to the energy gain of the electron per turn. In more modern synchrotrons, the energy loss is 300 million electronvolts per turn (Jackson, p. 667f.). And quantum effects can certainly be ignored here—in fact Planck's constant is of the order of $10^{-15}$ electronvolt seconds! Such considerations are made clear in problem 16.2 of Jackson's book (p. 769), where an equation is derived for the final orbital radius of an accelerating particle given the initial orbital radius—we have a non-negligible self-force in a non-quantum context.

---

[18] Belot's reason to keep things non-relativistic doesn't seem overly motivating. He wants to introduce rigid bodies, which are inconsistent with SR, but he's already said that electron models are not part of the theory, so this wouldn't be a *theory internal* inconsistency at least.

One option for Frisch is for him to renege on his claim that his conclusions extend to *relativistic* CED. He might claim that relativistic phenomena such as those associated with synchrotrons are not phenomena to be explained by the theory. Going non-relativistic would then also mean that he could criticise Belot's LFE, since Belot's LFE doesn't follow in the non-relativistic theory.[19] On the negative side Frisch would be required to make some significant changes to his book, since he draws on relativity on several other occasions. At any rate, the cumulative effects of the radiation would still stand.

Another possibility is to claim that phenomena where self-force effects become significant are so rare that they don't impinge on the theory he is delineating. Before synchrotron accelerators were built there really were no (known) phenomena where the self-force was significant. Frisch might try to account for such phenomena by drawing on what Batterman ([1995]) has called a 'theory between theories'. Just as certain phenomena 'between' the classical and the quantum are in practice handled by what Batterman calls 'semi-classical mechanics', synchrotron phenomena are handled by something which, strictly speaking, is not CED. After all, a different form of the LFE (usually the Abraham-Lorentz equation) is used to evaluate self-force effects in synchrotrons than in virtually all other applications of the theory. However, this isn't particularly satisfying, since Frisch cannot straightforwardly draw on Batterman. CED with the Abraham-Lorentz equation cannot be a theory between two different theories—what would be the *other* theory? At any rate, having just one consistent theory is conceptually much more satisfying than having one inconsistent theory and another theory for synchrotrons.

A final possibility is for him to claim that he *can* account for such phenomena within his theory, by ignoring $\mathbf{F}_{self} = 0$ and working out the energy loss from the MEs. This takes us to the second and most important difficulty for Frisch's conceptual scheme, the role played by $\mathbf{F}_{self} = 0$.

(F2) Frisch can respond to the previous difficulty in the following way. One can work from the MEs to derive an equation (the Abraham-Lorentz equation) which takes into account self-force effects just as Jackson does (p. 747f.). Now clearly he will have derived a contradiction if he allows $\mathbf{F}_{self} = 0$ to persist, so he will have to drop it for the time being. But this is OK, since on Frisch's view, 'for a given system we use only a proper subset of the theory's equations to model its behaviour' (p. 40). This solves difficulty F1.

---

[19] We see that both authors take the opposite stance to relativity to the one which most favours their position! Belot should embrace relativity, and Frisch should reject it.

But this brings into the spotlight the role played by $\mathbf{F}_{self} = 0$. We must ask on what grounds it deserves a place *in* the theory, the 'core assumptions', whereas other assumptions used in other contexts do not deserve a place. All sorts of what are commonly called approximation and idealisation assumptions are used within CED for the purposes of model-building, but these apparently do not go into Frisch's 'core'. How can he distinguish what goes in and what does not in such a way that $\mathbf{F}_{self} = 0$ deserves a place, but these other assumptions do not?

Judging from quotations already given I came to the conclusion that on Frisch's account what constitutes a theory are those assumptions which are readily used in model-building for classical EM phenomena. How often does an assumption need to be used to be considered a part of the theory? If only as often as $\mathbf{F}_{self} = 0$ is used, then Frisch will surely have to admit a whole host of other approximation assumptions into his theory along with $\mathbf{F}_{self} = 0$. In my view, $\mathbf{F}_{self} = 0$ isn't used in the vast majority of model-building in CED, if at all. Even when one models the trajectories of particles in synchrotron accelerators, and ignores the self-force in the first step (cf. Frisch, p. 33), the ignoring can be done by just staying silent on $\mathbf{F}_{self}$. Better to insist that assumptions must play an essential and regular role in model-building to be considered a part of the theory, and thus to kick $\mathbf{F}_{self} = 0$ out.

It might be argued that, when constructing a model, just staying silent on an issue doesn't tell us that it can be ignored in the given context; it is necessary, the argument goes, to make this clear with an equation. But many other things can also be ignored, and we don't write *their* absence into the theory. And at any rate, it can be worked out from the rest of the theory just how negligible the self-force is, as Jackson does (p. 746f.); the theory *already* tells us whether $\mathbf{F}_{self}$ is negligible in a given context. Whatever role $\mathbf{F}_{self} = 0$ is playing seems to me to be theory$_F$-external.

A further more general difficulty follows here. Since on Frisch's account 'for a given system we use only a proper subset of the theory's equations to model its behaviour' it follows that not all equations are used for all modelling purposes. But then how often does an equation need to be used to deserve a place in the 'core' of the theory? $\mathbf{F}_{self} = 0$ is a relatively easy candidate: we can say that it definitely doesn't deserve a place since it is never used. But other assumptions will be used occasionally, and it seems to follow that there will be no clear boundary between the 'core' assumptions and the 'non-core' assumptions. As we saw earlier, the account I attributed to Belot suffered from a similar difficulty.

Part of Frisch's argument for presenting his version of CED as the canonical version is that 'there is no satisfactory complete and consistent theory which governs classical phenomena involving charged particles.' (p. 26). However, it should by now be clear that there is at least a consistent possibility, and one that in fact includes only assumptions which Frisch *already* commits to. The solution is simply to drop $\mathbf{F}_{self} = 0$ from his theory, leaving a purely external LFE, $\mathbf{F}_{ext} = q(\mathbf{E}_{ext} + \mathbf{v} \times \mathbf{B}_{ext})$. Call this 'minimal CED'. Since $\mathbf{F}_{self} = 0$ was causing all of the difficulties for Frisch, this should be quite satisfactory. All the models he could derive with his theory$_F$ he can also derive with the 'minimal' theory. And this makes sense both within a relativistic and a non-relativistic setting.

Belot, we saw, was justified in sticking with his total-LFE in a *relativistic* setting, since then it follows from the external-LFE. However, Belot could also be satisfied with minimal CED, since if the total-LFE follows from the external-LFE the only difference between the two accounts of the theory is that minimal CED is more economical on axioms. The *content* of both theories will be the same, and which LFE we call *the* LFE will just be a terminological issue. In the *non*-relativistic setting we saw that Belot's LFE *didn't* follow from the rest of the theory. In this case, once again, he might stick with minimal CED. In this way he doesn't overstretch his commitments, and what *does* follow is then properly justified. This seems to suit his philosophy.

Therefore, I suggest, minimal CED can satisfy both authors, and in both a relativistic setting (with the necessary additions) and a non-relativistic setting. However, I stress that the *theory* is not the same for both authors. Whereas Belot would no doubt identify the theory with the axioms *and their closure*, Frisch would not. However, at least if the *presentation* of the theory is the same in both cases the possibility of miscommunication is reduced. The path is then clear for an analysis of the differences between these two 'philosophies of theory' in terms of the different methods of reasoning adopted.

## 6 Conclusion

At the end of §3 I considered two claims Frisch might be making: on the one hand that his theory$_F$ *is* CED, and on the other hand that it is an interesting and/or important unit of analysis for philosophy of science. Even the second, weaker claim has not stood up under analysis. The unit of analysis Frisch presents is conceptually most unsatisfying, in particular because there doesn't seem to be a principled reason why $\mathbf{F}_{self} = 0$ goes into the theory, and other candidate assumptions do not. Frisch's focus is still less important, since what I have called 'minimal CED' presents a unit of analysis which ought to satisfy all parties involved.

It isn't clear that this approach could be adopted generally, since the two different philosophies in play might give rise to different units of analysis on another occasion. However, it isn't such a coincidence that both parties can agree on the 'core' of the theory, since what the Frischian takes to be good for model-building, the Belotian identifies as a candidate for some species of doxastic commitment. Differentiating the positions in terms of reasoning styles may yet prove a valuable direction for research.

One outstanding difficulty is that of defining what constitutes the content of a given theory. We saw that on both Belot's and Frisch's account the boundary of a theory is not well-defined, and a cut-off such as Lakatos's hard core-auxiliary divide does not seem representative. This issue strikes me as one of particular importance for philosophy of science. If an account such as that of Frisch or Belot is essentially correct, then we need to dispense with the idea that the content of a theory has a well-defined boundary. This would be a fundamental change of perspective, and would strike to the heart of many analyses in the literature where the 'theory' plays a central role. With further research in this area, Fred Suppe's metaphor of the theory as the 'keystone' of philosophy of science ([1989], p. 429) may well turn out to be alarmingly apt.

## Acknowledgements

Many thanks to Steven French, Chris Timpson, Gordon Belot and Mathias Frisch for invaluable discussion and criticism.

*Division of History and Philosophy of Science*
*Department of Philosophy*
*University of Leeds*
*Leeds LS2 9JT,*
*UK*
*phl4pv@leeds.ac.uk*

## References

Batterman, R. [1995]: 'Theories Between Theories: Asymptotic Limiting Intertheoretic Relations', *Synthese*, **103**, pp. 171-201.
Belot, G. [2007]: 'Is Classical Electrodynamics an Inconsistent Theory?', *Canadian Journal of Philosophy*, **37**, pp. 263-82
Duffin, W. J. [1990]: *Electricity and Magnetism*, 4th edition, Maidenhead: McGraw-Hill.
Feynman, R., Leighton, R. and Sands, M. [1964]: *The Feynman Lectures on Physics*, vol. II, Reading, MA: Addison-Wesley.
Frisch, M. [2005]: *Inconsistency, Asymmetry, and Non-Locality*, Oxford: Oxford University Press.
Griffiths, D. [1999]: *Introduction to Electrodynamics*, Upper Saddle River, NJ: Prentice-Hall.
Jackson, J. [1999]: *Classical Electrodynamics*, 3rd edition, New York: John Wiley and Sons.

Kenat, R. [1987]: *Physical Interpretation: Eddington, Idealization and Stellar Structure Theory*, Ph.D thesis, University of Maryland.
Meheus, J. (*ed.*) [2002]: *Inconsistency in Science*, Dordrecht, The Netherlands: Kluwer Academic Publishers.
Muller, F. [2007]: 'Inconsistency in Classical Electrodynamics?', *Philosophy of Science*, **74**, pp. 253-77.
Rohrlich, F. and Hardin, L. [1983]: 'Established Theories', *Philosophy of Science*, **50**, pp. 603-17.
Suppe, F. [1989]: *The Semantic Conception of Theories and Scientific Realism*, Illinois: University of Illinois Press.